\documentclass[12pt]{article}

\begin{document}


\newcommand{\bb}{\begin{equation}}
\newcommand{\ee}{\end{equation}}
\newcommand{\bbb}{\begin{eqnarray}}
\newcommand{\eee}{\end{eqnarray}}
\newcommand{\vc}[1]{\mbox{$\vec{{\bf #1}}$}}
\newcommand{\mc}[1]{\mathcal{#1}}
\newcommand{\del}{\partial}
\newcommand{\lk}{\left}
\newcommand{\ave}[1]{\mbox{$\langle{#1}\rangle$}}
\newcommand{\re}{\right}
\newcommand{\pd}[1]{\frac{\del}{\del #1}}
\newcommand{\pdd}[2]{\frac{\del^2}{\del #1 \del #2}}
\newcommand{\Dd}[1]{\frac{d}{d #1}}
\newcommand{\sech}{\mbox{sech}}
\newcommand{\pref}[1]{(\ref{#1})}

\newcommand
{\sect}[1]{\vspace{20pt}{\LARGE}\noindent
{\bf #1:}}
\newcommand
{\subsect}[1]{\vspace{20pt}\hspace*{10pt}{\Large{$\bullet$}}\mbox{ }
{\bf #1}}
\newcommand
{\subsubsect}[1]{\hspace*{20pt}{\large{$\bullet$}}\mbox{ }
{\bf #1}}

\def\ie{{\it i.e.}}
\def\eg{{\it e.g.}}
\def\cf{{\it c.f.}}
\def\etal{{\it et.al.}}
\def\etc{{\it etc.}}

\def\AA{{\cal A}}
\def\BB{{\cal B}}
\def\CC{{\cal C}}
\def\DD{{\cal D}}
\def\EE{{\cal E}}
\def\FF{{\cal F}}
\def\GG{{\cal G}}
\def\HH{{\cal H}}
\def\II{{\cal I}}
\def\JJ{{\cal J}}
\def\KK{{\cal K}}
\def\LL{{\cal L}}
\def\MM{{\cal M}}
\def\NN{{\cal N}}
\def\OO{{\cal O}}
\def\PP{{\cal P}}
\def\QQ{{\cal Q}}
\def\RR{{\cal R}}
\def\SS{{\cal S}}
\def\TT{{\cal T}}
\def\UU{{\cal U}}
\def\VV{{\cal V}}
\def\WW{{\cal W}}
\def\XX{{\cal X}}
\def\YY{{\cal Y}}
\def\ZZ{{\cal Z}}

\def\sinh{{\rm sinh}}
\def\cosh{{\rm cosh}}
\def\tanh{{\rm tanh}}
\def\sgn{{\rm sgn}}
\def\det{{\rm det}}
\def\trace{{\rm Tr}}
\def\exp{{\rm exp}}
\def\sh{{\rm sh}}
\def\ch{{\rm ch}}

\def\ell{{\it l}}
\def\str{{\it str}}
\def\lp{\ell_{{\rm pl}}}
\def\blp{\overline{\ell}_{{\rm pl}}}
\def\ls{\ell_{{\str}}}
\def\bls{{\bar\ell}_{{\str}}}
\def\bM{{\overline{\rm M}}}
\def\gs{g_\str}
\def\gym{{g_{Y}}}
\def\geff{g_{\rm eff}}
\def\eff{{\rm eff}}
\def\r11{R_{11}}
\def\kel{{2\kappa_{11}^2}}
\def\kten{{2\kappa_{10}^2}}
\def\lpten{{\lp^{(10)}}}
\def\alp{{\alpha'}}
\def\aleff{{\alp_{eff}}}
\def\sqaleff{{\alp_{eff}^2}}
\def\tgs{{\tilde{g}_s}}
\def\talp{{{\tilde{\alpha}}'}}
\def\tlp{{\tilde{\ell}_{{\rm pl}}}}
\def\tr11{{\tilde{R}_{11}}}
\def\wtilde{\widetilde}
\def\what{\widehat}
\def\hlp{{\hat{\ell}_{{\rm pl}}}}
\def\hr11{{\hat{R}_{11}}}
\def\hf{{\textstyle\frac12}}
\def\coeff#1#2{{\textstyle{#1\over#2}}}
\def\CY{Calabi-Yau}
\def\lessapprox{\;{\buildrel{<}\over{\scriptstyle\sim}}\;}
\def\greaterapprox{\;{\buildrel{>}\over{\scriptstyle\sim}}\;}
\def\inbar{\,\vrule height1.5ex width.4pt depth0pt}
\def\IC{\relax\hbox{$\inbar\kern-.3em{\rm C}$}}
\def\IR{\relax{\rm I\kern-.18em R}}
\def\IP{\relax{\rm I\kern-.18em P}}
\def\Z{{\bf Z}}
\def\R{{\bf R}}
\def\One{{1\hskip -3pt {\rm l}}}
\def\sst{\scriptscriptstyle}
\def\osc{{\rm\sst osc}}
\def\lam{\lambda}
\def\lc{{\sst LC}}
\def\pr{{\sst \rm pr}}
\def\cl{{\sst \rm cl}}
\def\D{{\sst D}}
\def\bh{{\sst BH}}
\def\vev#1{\langle#1\rangle}

\def\alp{{\alpha}} 

\input{epsf}

\begin{titlepage}
\rightline{CLNS 99/1657}

\rightline{hep-th/0002126}

\vskip 1cm
\begin{center}
\Large{{\bf 
Comments on D branes\\ and the renormalization group
}}
\end{center}

\vskip 1cm
\begin{center}
Vatche Sahakian\footnote{\texttt{vvs@mail.lns.cornell.edu}}
\end{center}
\vskip 12pt
\centerline{\sl Laboratory of Nuclear Studies}
\centerline{\sl Cornell University}
\centerline{\sl Ithaca, NY 14853, USA}

\vskip 2cm

\begin{abstract}
We review the de Boer-Verlinde-Verlinde formalism for the renormalization
group in the context of Dp brane vacua. We comment on
various aspects of 
the dictionary between bulk and boundary and relate
the discussion to the Randall-Sundrum scenario.
We find that the gravitational coupling for the Randall-Sundrum gravity 
on the Dp brane worldvolume is proportional to the central charge 
of the Yang-Mills theory. We compute the beta function
and find the expected uneventful flow prescribed
by the classical dimension of the Yang-Mills operator. Finally, we
argue for a dynamical mechanism for determining the cosmology on
the brane.
\end{abstract}

\end{titlepage}
\newpage
\setcounter{page}{1}

\newcommand{\nsection}[1]{{\vspace{15pt}\large{\bf #1\ : }}\vspace{12pt}}

\section{Introduction}

At the foundation of the string theoretical incarnation of the
holographic principle~\cite{MALDA1,WITHOLO,KLEB,MALDA2}
is an intriguing relation between 
renormalization group (RG) flow and the equations of motion of gravity.
This connection was made recently more
transparent in the work of~\cite{dBVV}
by casting the Einstein equations into the form of Hamiltonian
evolution across timelike foliations.  In this note, 
we present a series of comments regarding this approach
in the context of studying RG flow in
Dp brane geometries.

The proposal~\cite{dBVV}
raises many intriguing questions regarding
the dictionary between bulk and boundary. We list a sample
of questions that drove us to investigate this subject:
\begin{itemize}
\item Given that the prescription involves an
explicit interpretation of
the gravitational Hamiltonian equations as describing RG
evolution, one may wonder whether 
this approach amounts to reconstructing the boundary theory within
the gravitational initial value problem. There is then an apparent
mismatch between the amount of data required to define a boundary theory and
that required to uniquely specify the bulk.

\item One would like to clarify  
the proper interpretation of the Randall-Sundrum gravity~\cite{RSGRAV}
on the boundary in the context of the flow equations. This has to do 
with studying the bulk/boundary correspondence with a 
{\em finite} cutoff and understanding the meaning of the slicing of a space
from the perspective of renormalization of the boundary theory.

\item The principle of renormalization group~\cite{WILSON} 
prescribes the general behavior
of a quantum field theory as a function
of energy scale. As a statement in its most fundamental form, 
it relates the theory
at different energy scales through diffeomorphisms in the space
of its couplings. It is known however that
this flow may also be endowed
with additional structures~\cite{ZAMO,WZ,LASSIG,JO}. 
Of particular physical interest
is a possible mechanism for driving it by
a scalar function, the so-called c-function~\cite{ZAMO,CFLT}.
Beta functions of a theory for example may be measures of the gradients of 
this function in coupling space. In this context,
we expect that such additional features of
RG flow play an important role in its relation
to gravitational physics. One would like to determine
how much structure the bulk
equations impose on the boundary theory.

\item There is room to better understand the dynamical aspects of
cosmology in the Randall-Sundrum gravity~\cite{KRAUSCOSMO,GUBSERRS,TYECOSMO} 
from the bulk point of view; the tuning of the
brane tension so as to cancel the cosmological constant must be determined
by the physics.
Unravelling the content of the canonical
momentum variable of the induced metric may provide part of the answer.

\item The Hamiltonian evolution equations do not care about a 
preferred direction or ``arrow''
for the flow; whereas renormalization group does. Introducing this
arrow into the formalism is a highly non-trivial statement related to
a principle introduced by Bousso~\cite{BOUSSOCONJ,BOUSSOCONJ2}.
\end{itemize}

We will attempt to comment on these issues in the process of examining
extremal and near extremal Dp brane geometries. Let us
mention as well that recently an interesting tangential 
exploration of some of these issues
were carried out in an attempt to find
a relation directly between closed and open string theories~\cite{KV}.
Related interesting work also appeared in~\cite{SCHMID1} and~\cite{NOZ}.
The preprint~\cite{GKR} contains overlapping
and complimentary material to this work.

We present our discussion as a review of~\cite{dBVV} 
generalized to arbitrary dimensions, inter-dispersing the text with
our examples and comments.

\section{The de Boer-Verlinde-Verlinde formalism}

We consider a bulk gravitational theory in $D$ space-time dimensions,
with matter added in the form of an arbitrary number of
scalar fields~\cite{dBVV}
\bb\label{action}
S_D=\alp_G \int_D \lk( R^{(D)}+2\Lambda\re)
- \alp_M \int_D 
\frac{1}{2} G_{IJ}(\phi) \nabla_\mu \phi^I \nabla^\mu \phi^J - V(\phi)\ .
\ee
We assume we are given some
vacuum of this action corresponding
to a trajectory (or a section of a trajectory) of renormalization group flow
of a $D-1$ dimensional ``boundary theory''. We embed an
arbitrary {\em timelike} surface in this space 
and choose Gaussian normal coordinates
with respect to it (see the appendix for definitions and
conventions used). We then cut the space along this surface while
focusing on excitations in the region bounded by the cut. There will
generally be two such regions, and 
the criterion is to pick the one through
which null geodesics projected from the boundary tend to converge. 
This statement was first
proposed by Bousso in the context of entropy bounds on 
gravitational vacua~\cite{BOUSSOCONJ,BOUSSOCONJ2}.
The convergence of the geodesics is a statement related to
the monotonicity of the c-function of the boundary theory~\cite{CFUNCTION}; 
regions of spacetime away from the boundary and 
deep into the bulk are then to be associated with
physics at lower energy scales.
Slicing space in this manner can be done
at the expense of introducing a boundary term to the action
\bb\label{totaction}
S_{tot}=S_D + 2\alp_G \int_{D-1} K\ .
\ee
$K_c^c\equiv K$ is the trace of the extrinsic curvature of the boundary
\footnote{Our definition for the normals to the boundary is such that
they point {\em inward} toward the bulk.}. We then have
\bb\label{var}
\frac{\delta S_{tot}}{\delta h^{ab}}
= \alp_G \sqrt{h} \lk(K_{ab}-h_{ab} K\re)\equiv \alp_G \sqrt{h}\ \tau_{ab}\ ,
\ee
where $h_{ab}$ is the induced 
metric on the boundary. The additional piece assures that vacua
of~\pref{action} are at minima of~\pref{totaction} when perturbations
of the metric vanish at the boundary.
The extrinsic curvature $K_{ab}$ can be 
viewed from the perspective of the $D$ dimensional bulk 
as an energy-momentum tensor with delta function support
on the boundary.
The space outside this region is thrown away and the interface at the cut
may be thought of as sourcing the bulk. One is also left with boundary 
conditions on the scalar fields. In parallel to~\pref{var}, we also define
\bb\label{scalardef}
\Omega_I\equiv \frac{1}{\alp_M \sqrt{h}} \frac{\delta S}{\delta \phi^I}\ .
\ee

This setup is a low energy approximation of string theory with an IR
cutoff; it is proposed that this string theory 
has a dual description through a $D-1$ dimensional
effective boundary quantum field theory
with a UV cutoff.  The partition functions of the
bulk and boundary theories are to be equated; in the low energy
supergravity approximation, we have 
then the statement~\cite{WITHOLO,KLEB,dBVV}
\bb\label{sz}
i\ S=\ln\ Z_{bndry}^{(\mu)}\ ,
\ee
where $\mu$ is the UV cutoff scale, related to the location of
the cut, and the dual theory is naturally coupled to
the metric $h_{ab}$. In this approximation, we are
treating the latter as a classical background 
to the boundary quantum field theory. 

The forms of equations~\pref{var} and~\pref{scalardef} suggest writing
the bulk equations of motion in the
Hamiltonian formalism.
Hamiltonian flow from the boundary 
is a constrained system,
as is typical of theories with gauge
symmetries. The choice of the cut is arbitrary, and,
fixing one,
the system is still endowed with redundancies; these are handled with
two sets of constraint equations on the initial value data at the boundary.
The first is a statement regarding $D-1$ dimensional Poincar\'{e}
invariance
\bb\label{encons}
D^a \tau_{ab}=\frac{\alp_M}{2\alp_G} \Omega_J D_b \phi^J\ ;
\ee
the second and the more interesting one is less transparent~\cite{dBVV}
\bbb\label{cons}
&\mbox{ }&\alp_G \lk( R+2\Lambda\re) + \alp_G \lk(\tau_{ab} \tau^{ab}
-\frac{\lk(\tau_c^c\re)^2}{D-2}\re) \nonumber \\
&-&\alp_M \lk(\frac{1}{2} G_{IJ} D_a\phi^I D^a \phi^J
-\frac{1}{2} G^{IJ} \Omega_I \Omega_J\re)+\alp_M V=0\ ,
\eee
The bulk/boundary correspondence proposes to replace the bulk action
appearing in these equations through $\Omega_I$ and $\tau_{ab}$
with that of an effective boundary 
theory in $D-1$ dimensions. Equation~\pref{cons} is one of the
most important statements of~\cite{dBVV}. Recently, its relevance
in perturbative regimes was also investigated~\cite{KV}.

Typically, renormalization group flow
of a quantum field theory coupled to a classical background metric
induces an Einstein gravity term in the effective action.
A computation of the expectation value for the quantum field theory 
energy momentum tensor renormalizes both the Einstein tensor and 
the cosmological constant~\cite{BANDD}. 
Therefore, a general form for the effective
action at scale $\mu$ is given by~\cite{dBVV}
\bb\label{baction}
S_{bndry}=\int_{D-1} \sqrt{h} \lk( \Phi(\phi) R + U(\phi)
+\frac{1}{2} M_{IJ}(\phi) D_a\phi^I D^a\phi^J\re)
+\Gamma_{eff}^\mu
+\cdots\ .
\ee
$R$ denotes the scalar curvature constructed from the
metric $h_{ab}$; we will write $D^a$ for the covariant derivatives
associated with this metric with the latin indices running over the 
$D-1$ boundary coordinates. $\Gamma_{eff}^\mu$ is the effective
action of the boundary theory at scale $\mu$ with
\bb
\frac{1}{\sqrt{h}}
\frac{\del \Gamma_{eff}^\mu}{\del \phi^I}\equiv \lk<\OO_I\re>\ ;\ \ \ \ 
\frac{1}{\sqrt{h}}
\frac{\del \Gamma_{eff}^\mu}{\del h^{ab}}\equiv 
\lk<t_{ab}\re>\propto \lk< T_{ab} \re>\ ,
\ee
the $\OO_I$'s being the operators appearing in the boundary theory, and
$\lk< T_{ab} \re>$ being the part of the theory's energy momentum tensor 
that, by definition, acquires only the beta function anomaly
\bb\label{tcc}
\lk<T^c_c\re>=\beta^I \lk<\OO_I\re>\ .
\ee
The normalization $2 t_{ab}=T_{ab}$ will be determined from
the Hamiltonian equations later.
The boundary values for the scalars $\phi^I$ are to be equated with
the {\em dimensionless} coupling constants of the boundary theory
as measured at the scale $\mu$. We have absorbed any explicit
appearances of $\mu$
into the definition of the $\lk<O_I\re>$'s. The effective action is
an infinite sum where all operators and higher derivative terms of
the background metric are allowed subject to the symmetries
inherited from the bulk. 
The $\Phi\ R$ term is referred to in the literature as the Randall-Sundrum
gravity~\cite{RSGRAV}. In this context, it acquires dual
interpretations: from the point of view of the bulk, 
it is the zero mode of the
$D$ dimensional graviton dynamically confined to the boundary.
From the boundary point of view, it is generated at low energies
by the flow of the boundary quantum field
theory. Note that the latter is 
part of a holographic image of the $D$ dimensional gravitational bulk. 
Both
viewpoints then eventually amount to the same physical interpretation
for the origin of the Randall-Sundrum gravity.

It is easy to see that the form of equation \pref{cons}
determines the coefficients of all the local terms in the boundary action.
One treats the metric $h_{ab}$ and the scalars $\phi^I$ as arbitrary
classical fields thus generating a set of relations for the
unknown functions $U$, $\Phi$ and $M_{IJ}$ of~\pref{baction} 
\bb\label{c1}
2\alp_G^2 \Lambda +\alp_M \alp_G V=
\frac{D-1}{4(D-2)} U^2
-\frac{1}{2} \frac{\alp_G}{\alp_M}G^{IJ} \del_I U \del_J U\ ;
\ee
\bb\label{c2}
\alp_G^2= \Phi U \frac{D-3}{2 (D-2)}
-\frac{\alp_G}{\alp_M} G^{IJ} \del_I \Phi \del_J U\ ;
\ee
\bb\label{c3}
-\beta^K \lk(\del_K M_{IJ}-\del_I M_{KJ}-\del_J M_{KI}\re)=
2 (D-2) \frac{\alp_G \alp_M}{U} G_{IJ}+(D-3)M_{IJ}\ .
\ee
One is also left, to this order in the expansion, with expressions
originating from $\Gamma_{eff}^\mu$
\bb\label{tcc2}
\lk< t^c_c\re> = - \lk(\frac{\alp_G}{\alp_M}\re) \frac{D-2}{U}
G^{IJ} \del_I U
\lk<\OO_J\re>\ ,
\ee
and
\bb
\lk<t_{ab}\re> \lk< t^{ab} \re> =
\frac{1}{D-2} \lk<t^c_c\re>^2
-\frac{\alp_G}{2\alp_M} G^{IJ} \lk<\OO_I\re> \lk<\OO_J\re>\ .
\ee

Relations involving non-local data about the boundary theory
arise from the first order differential equations
describing the evolution of the system;
we will later on explore this additional  information in some detail. 
For now, let us apply
what we have up to now to AdS and Dp brane geometries.

\section{Gravity on the worldvolume of Dp branes}

We take the AdS metric in $D$ dimensions as
\bb
ds^2=\frac{u^2}{l^2} \lk(dx_{D-1}^2\re) + \frac{l^2}{u^2} du^2\ .
\ee
We choose the foliation $u=\mbox{constant}$, and cut the space along $u=u_c$. 
The space $u<u_c$ is candidate for holography; the $D-1$ dimensional
boundary theory is conformal with a UV cutoff $u_c$ breaking this symmetry
in a trivial manner.
Setting $V=0$ and $D^a \phi^I=0$ in~\pref{action},
we obtain from~\pref{c1} and \pref{c2}\footnote{
The sign of $U$ cannot be determined from~\pref{c1}; but once its sign
is fixed, that of $\Phi$ follows. We have determined these signs by required
positivity of the boundary theory's energy momentum tensor; this will
become clear in equation~\pref{enbal}.}
\bb
U=-\alp_G \sqrt{8\Lambda \frac{D-2}{D-1}}\ ,\ \ \ 
\Phi=-\alp_G \sqrt{\frac{(D-1)(D-2)}{2\Lambda (D-3)^2}}\ .
\ee
These relations were also found in~\cite{BALASTRESS} 
by requiring the finiteness of
the energy content of the bulk vacuum as the boundary is taken to infinity.
In the renormalization group language, the method corresponds to adding
counterterms so as to cancel divergences resulting from taking the UV
cutoff to infinity. The boundary action then takes the form
\bb\label{bactionads}
-S_{bndry}=\frac{\alp_G^{(D-2)} l}{D-3} \int_{D-1} \sqrt{h} 
\lk(R+2 \Lambda^{(D-1)}\re)+\cdots \ ,
\ee
with the $D-1$ dimensional parameters
\bb\label{param}
\alp_G^{(D-1)}=\frac{\alp_G l}{D-3} (\mu l)^{D-3}\ ,\ \ 
2\Lambda^{(D-1)}=2 \frac{(D-2)(D-3)}{l^2} (\mu l)^2\ ;
\ee
The scale $l$ is defined by $ 2\Lambda\equiv \frac{(D-1)(D-2)}{l^2}$,
which, in our convention, is positive for AdS spaces.
From the UV-IR relation~\cite{PEETPOLCH}, we have used $\mu\sim u_c/l^2$. 
The UV cutoff scale $\mu$ appears in these equations since
we have rescaled the metric so as to make $h_{ab}$
Minkowski. This is because the boundary theory corresponding
to our choice of foliation
lives in flat space. More on this issue later. 
In the literature, the space is often arbitrarily cut at
the energy scale $\mu=1/l$, and we recognize 
the Randall-Sundrum relation~\cite{RSGRAV}
$G_4=2 G_5/l$ for $AdS_5$ ($\alp_G\equiv 1/16\pi G_D$).
More generally, collecting the factors of $l$ together in \pref{param},
we are fixing the value of the
dimensionless gravitational coupling and the cosmological constant
as measured at a cutoff scale $\mu$. Note also that all
terms appearing in~\pref{bactionads}
remain finite in the decoupling limit (as we shall also see
explicitly later on).

Along the same line of thought,
one may look
at $R^2$ corrections to this boundary gravity. These would also be
interpreted as being generated by the flow of the boundary quantum
field theory to lower energy scales. Adding to the action the terms
\bb\label{rsquare}
-S_{bndry}\rightarrow \int_{D-1} \sqrt{h}
\lk( A(\phi) R^2 + B(\phi) R_{ab} R^{ab} + C(\phi) R_{abcd} R^{abcd} \re)\ ,
\ee
and substituting this in the constraint equation~\pref{cons}, 
one obtains, in the case where the boundary theory is conformal,
\bb
A=-\frac{\alp_G}{4} \frac{(D-1)}{(D-3)^2 (D-5) (D-2)} l^3\ ,\ \ \ 
B= \alp_G \frac{l^3}{(D-3)^2 (D-5)}\ ,\ \ \ 
C=0\ .
\ee
These are precisely the expressions proposed in~\cite{EJM,NOOFINITE} 
as the counterterms
needed to keep the action finite when the boundary is sent to
infinity\footnote{In comparing the coefficient, note our normalization of
the Einstein term in~\pref{baction}.}. 
We note that the pole for $D=5$ in these expression is harmless;
as shown in~\cite{EJM}, when evaluated for the AdS vacuum configuration, these
factors cancel. The duality between bulk and boundary would be one
expanded about this vacuum configuration, so that we expect these
terms to remain as finite corrections to the $D-1$ dimensional Einstein
gravity. Rescaling the boundary metric so as to make it
Minkowski for the AdS vacuum,  we see that the $R^2$ terms scale as
$R^2 l^2 (\mu l)^{-2}$; and with
$\mu_0$ being the energy scale of a process
under consideration in the boundary theory, an infinite
expansion in higher derivative of the metric will correspond to an expansion
in powers of $\mu_0/\mu$ ($\mu$ being the cutoff energy).
Therefore, these corrections become important as we study processes
in the boundary theory with energy progressively closer
to the cutoff scale. This effect may be related to observations made
in~\cite{SUSSHOLO,BALARG} 
regarding non-local effects in the holographic image of the bulk; 
we will talk more on this issue in the next section.

While in the subject of higher order corrections to the Randall-Sundrum
gravity, let us qualitatively analyze the effects of $R^2$ corrections
to the {\em bulk} action; the physical origin of such corrections being 
expansions in the string tension and the string coupling.
The $D$ dimensional scalar curvature $R^{(D)}$
splits, as outlined in the appendix, into the $D-1$ dimensional scalar
curvature $R$ and the extrinsic curvature of the boundary $K$,
schematically, as
$R^{(D)}\sim R+K^2$. For corrections to Einstein gravity in the bulk
of the form $R^\alpha$, where $\alpha$ is a positive integer,
we induce in the Lagrangian, among a series
of other corrections, a term of the form $K^{2\alpha}$. This means
that the canonical momentum $\tau$ will appear in the expression for $K$ as
$K\sim \tau^{2\alpha -1}$, amongst an infinite sequence of powers of 
$\tau$ and $R$. The Hamiltonian will then involve a term of the form 
$\tau^{2\alpha}$, with essentially the same coefficient as in the original
$R^2$ term of the bulk action. Tracing carefully the lapse
function in this process, one finds that the constraint equation~\pref{cons}
will get corrections involving powers of $\delta S/\delta h^{ab}$ and $R$; 
in particular, the term we traced above leads to a contribution of the
form
\bb\label{corr}
\rightarrow\lk(\frac{\delta S}{\delta h^{ab}}\re)^{2\alpha}\ .
\ee
In string theories with 32 supersymmetries, the first corrections will
have $\alpha=4$, and the coefficient will be proportional to
${\alpha'}^3$. This implies that the gravitational coupling of the
Randall-Sundrum gravity will in general get stringy corrections.
We will argue later on that
this coupling is also expected to be proportional to
the c-function of the boundary theory.
For Dp brane geometries with $p\neq 3$, 
we will see that the first non-trivial contribution
to the Yang-Mills beta function comes from such contributions
\footnote{
In the case of D3 branes, 
it then better be the case that 
the coefficients of the scalar curvature arising from
terms such as~\pref{corr} sum to zero; for otherwise
the central charge of the conformal
Yang-Mills theory would get corrections of order
$(\gym^2 N)^{-3/2}$ for $\alpha=4$. 
Perhaps this is related to the observation that 
the $AdS_5\times S^5$ solution does not get modified in the presence
of higher order string corrections as shown in~\cite{KALLR}.}.
Finally, let us note that
the $R^2$ terms in the boundary action that we just discussed 
will generally
mix with terms of similar form arising from our discussion
in the previous paragraph. There may however
be a hierarchy between these two sets of contributions if we assume that
operators in
the boundary theory coupling to the ``tails'' of stringy states 
in the bulk will have
much higher mass dimension than those 
responsible for the $R^2$ terms
of the previous discussion.

We next look for the Randall-Sundrum gravity for the
case of Dp brane geometries; \ie\ the classical gravitational sector 
coupled to $p+1$ dimensional Super Yang-Mills (SYM) theory;
the extremal
Dp brane solution is used in the bulk and
corresponds to a canonical choice for the SYM vacuum in
Minkowski space. To apply the previous formalism to this
case, we would need to do a little of juggling to put the supergravity
action in the desired form~\pref{action}.
This can easily be achieved
by using the results of~\cite{BST,CO}. One starts from IIA or IIB theory
in the string frame, and goes to the so called ``holographic dual frame''
\bb
g^{(dual)}_{ab}=
\lk(Ne^\phi\re)^{\frac{2}{p-7}} g^{(str)}_{ab} \ ,
\ee
where $\phi$ is the dilaton and $N$ is the number of Dp branes.
For extremal Dp brane vacua, we also have
\bb\label{dilaton}
Ne^\phi=\lk(\gym^2 N \lk(\frac{5-p}{2} u \re)^{p-3}\re)^{\frac{7-p}{2(5-p)}}
\equiv\lk(\geff^2\re)^{\frac{7-p}{2(5-p)}}\ ,
\ee
where $\gym^2=\gs \ls^{p-3}$, and $u$ appears in the dual holographic frame as
\bb\label{extremal}
ds_{dual}^2=\alpha' \lk(\frac{4 u^2}{(5-p)^2} \lk(-dt^2+dx_{(p)}^2\re)
+\frac{(5-p)^2}{4 u^2} du^2+d\Omega_{8-p}^2\ \re)\ .
\ee
We identify $u$ as the energy scale of the
SYM theory in flat $p+1$ dimensional space; and
$\geff^2$ is the effective dimensionless Yang-Mills coupling measured
at energy scale $u$. One then dimensionally reduces on
the transverse $8-p$ sphere of constant radius to $D=p+2$ dimensions.
Rescaling to the Einstein frame
\bb
g^{(dual)}_{ab}=
\lk(Ne^\phi\re)^{\frac{4}{p}\frac{p-3}{p-7}} g_{ab} \ ,
\ee
one gets the action
\bb\label{dwframe}
S_{Dp}=\frac{N^2\ \Omega_{8-p}}{(2\pi)^7 \ls^p} \int d^{p+2}x\ 
\sqrt{|g|} \lk( R-\frac{1}{2} \lk( \del \Pi\re)^2 +V\re)\ .
\ee
Here $\Pi$ is related to the dilaton field by
\bb
\Pi=\frac{2\sqrt{2(9-p)}}{\sqrt{p} (7-p)}\phi\ ,
\ee
and
\bb
V\equiv\frac{1}{2\alpha'} 
(9-p)(7-p) \lk(Ne^\phi\re)^{\frac{4}{p} \frac{p-3}{p-7}}\ .
\ee
We foliate the
$p+2$ dimensional space with $u=\mbox{constant}$ surfaces and
cut it at a certain $u=u_c$. The region $u<u_c$ has a dual description
through the effective $p+1$ dimensional SYM theory with UV cutoff $u_c$.
We then apply the previous formalism, and find
\bb\label{results}
U=-\frac{\alp_G (9-p)}{\ls}
\lk(Ne^\phi\re)^{\frac{2}{p} \frac{p-3}{p-7}}\ ,\ \ \ 
\Phi=-\frac{\alp_G}{2} \ls
\lk(Ne^\phi\re)^{-\frac{2}{p} \frac{p-3}{p-7}}\ ,
\ee
\bb\label{mgeff}
M_{\geff^2}=\ls \frac{\Omega_{8-p}}{(2\pi)^7} 
\frac{9-p}{p (5-p)^2}
\frac{c(\phi)}{\geff^4}\ .
\ee
Here, $M_{\geff^2}$ is the metric over the one dimensional Yang-Mills
coupling space; to obtain it, we have rescaled the boundary metric to
the holographic dual frame, and used the relation between 
$\Pi$ and $\geff^2$; \ie\ this term appears in the boundary theory as
$\int M_{\geff^2} \del_a \geff^2 \del^a \geff^2$. 
We have introduced the function $c(\phi)$
\bb
c(\phi )\equiv N^2 \lk( Ne^\phi\re)^{2\frac{p-3}{7-p}}\ .
\ee
The gravitational part of the boundary action then becomes
\bb\label{sdp}
S_{Dp}=\frac{\Omega_{8-p}}{2 (2\pi)^7 \ls^{p-1}} 
\lk(\frac{2 \ls}{5-p}u_c\re)^{p-1} c(\phi)
\int_{p+1}\ \sqrt{h}\ 
\lk(R
+2 \frac{(9-p)}{\ls^2}\lk( \frac{2 \ls}{5-p}u_c\re)^2
\re)+\cdots \ .
\ee
We have also rescaled the metric
to a flat Minkowski form. Now, if one uses~\pref{dilaton},
the function $c$ dressing the gravitational coupling gets identified
with the c-function
\bb\label{cfunc}
c=N^2 \lk(\geff^2\re)^{\frac{p-3}{5-p}}\ .
\ee
This expression was found in~\cite{CFUNCTION} using a proposal that relates
the c-function to the rate of acceleration of null geodesics
in the geometry. It was also shown to interpolate correctly between
the known asymptotics from string theory.
For the $AdS_5$ case, this dressing of $G_4$ is
well known. We see here that it generalizes for all Dp branes with $p<5$.
It is a statement which is consistent with our intuitive picture
that gravity would be induced by all of the degrees of freedom of the
holographic boundary theory.
The appearance of $u_c\sim \mu$ is similar in physical content to
the AdS cases above: The effective
gravitational coupling of the Randall-Sundrum gravity on the Dp branes
is given by the c-function of the Super-Yang Mills theory measured at
the cutoff scale $\mu$. 
Note also that all factors of 
the string tension cancel in~\pref{sdp} 
and all remains finite in the decoupling limit.
Finally, we note that the $p=5$ scenario is
problematic as the near horizon geometry is Minkowski space and the relation
between the bulk extent and energy scale on the boundary is less
understood. Technically, the conformal transformations we applied 
break down for $p=5$ as can be seen in the equations above. We hope
we will return to this case in the future.

\section{The initial value formulation}

In the Hamiltonian formulation of the dynamics, the canonical
variables and their momenta are are $h_{ab}$, $\phi^I$, $\tau_{ab}$ and
$\Omega_I$. The initial value problem for gravity assures the existence
of a bulk vacuum, unique up to diffeomorphisms, once the
boundary values of these variables are specified.
We will argue below that the need for this initial value data
adapts well into a renormalization group
interpretation. However, this will be only a partial 
understanding of the consistency of the
second order nature of the bulk equations of motion with renormalization
group flow and we will have to come back to this issue in the last 
section.

The roles of 
the canonical variables $h_{ab}$ and $\phi^I$ in the boundary theory
are straightforward to understand. $h_{ab}$ is the classical background
metric to which one couples the boundary quantum field theory, 
and the $\phi^I$'s map onto the couplings of this quantum field theory. 
In cases where one encounters relevant or irrelevant operators in the
$D-1$ dimensional effective theory, the boundary values for the
scalars are to be taken as the dimensionless couplings of these operators
measured at the cutoff scale, as we will show explicitly in the
Dp brane scenarios below. The interpretation of the canonical 
momenta is also straightforward when one writes them,
using equations~\pref{var}, \pref{scalardef} and~\pref{baction}, 
in terms of the boundary action parameters
\bbb\label{enbal}
\alp_G \lk(K_{ab} - h_{ab} K\re)&=&
\Phi G_{ab}-\frac{1}{2} h_{ab} U +\lk< t_{ab}\re> \nonumber \\
&+&\frac{1}{2} M_{IJ} D_a \phi^I D_b\phi^J
-\frac{1}{4} h_{ab} M_{IJ} D_c \phi^I D^c \phi^J  \nonumber \\
&+&h_{ab} \del_I \del_J \Phi D_c \phi^I D^c \phi^J+h_{ab} \del_I\Phi
D_c D^c\phi^I \nonumber \\
&-&\del_I \del_J \Phi D_a \phi^I D_b \phi^J
-\del_I \Phi D_a D_b \phi^I+ \cdots \ ,
\eee
\bb\label{opexp}
\alp_M \Omega_I= R \del_I \phi +\del_I U \lk< \OO_I\re>+
\frac{1}{2} \del_I M_{JK} D_a \phi^J D^a \phi^K
- D^a\lk( M_{IJ} D_a \phi^J\re)+\cdots \ .
\ee
A simpler form of equation~\pref{enbal}, with $D_a\phi^I=0$ 
was written in~\cite{BALASTRESS,GUBSERRS}. 
It is an energy balance equation between boundary and bulk/boundary interface.
Comments relating to equation~\pref{opexp} were eluded to
in~\cite{BALARG,BALABULK2,BFB}.
In the Hamiltonian formulation, these two statements are simply
the definitions of the canonical momenta at the boundary.
Specifying the initial values for $\tau_{ab}$ and $\Omega_I$
is necessary to determine a physically unique bulk solution, in addition to the
specification of the induced metric and the boundary values 
of the scalar fields. It may be expected that, to define the effective
dual boundary theory, all one needs is the classical background metric and
the dimensionless values of the couplings at the cutoff scale. One then
singles out a renormalization group trajectory in the space of couplings,
and hence a bulk solution describing physics at lower energy scales. It
is obvious however that such an interpretation 
is inconsistent with the bulk initial value formulation.
A proper resolution to this problem may be reached if we
slightly revise the statement of the bulk/boundary correspondence:
The latter
needs to be taken as an identification between two theories as each is
expanded about a ``common vacuum'' configuration; \ie\ one
needs also to specify a map between the corresponding states.
The boundary theory in general may be expanded about a unique vacuum,
or a finite temperature one, or one which is a choice out of many if the
theory is endowed with a moduli space (as in the case of
bulk vacua representing multicenter BPS configurations). In general,
it is then necessary to specify a map between bulk and boundary vacua.
Focusing on~\pref{opexp}, specifying the value of $\Omega_I$ in addition
to $h_{ab}$ and $\phi^I$ amounts to fixing the one point correlation
functions of the boundary operators, \ie\ one specifies the boundary vacuum
of interest. Fixing
$\tau_{ab}$ as well 
would amount to specifying the vacuum expectation 
value of the energy momentum tensor. 

However,
this is not the whole story. Let us focus on~\pref{enbal}.
Given a bulk solution, we should expect that 
cutting the space along any arbitrary timelike surface does not
affect the energy momentum tensor's vev, except for a redshift
factor arising from the different choice of time.
From the point of view of the boundary theory,
this phenomenon is highly non-trivial; we will illustrate this statement 
through the example of near extremal D3 branes.
We take the metric as
\bb
ds^2=\frac{u^2}{l^2} \lk( -h(u) dt^2 + dx_{(3)}^2\re) 
+ \frac{l^2}{u^2} h(u)^{-1} du^2\ .
\ee
$l=\ls (\gs N)^{1/4}$ is the AdS scale, $h(u)=1-(u_0/u)^4$ with
$u_0=c_1 \sqrt{\gs N} T$ where $T$ is the Hawking temperature and $c_1$
is a numerical constant. For all cuts of this space we consider,
we define the coordinate $\tau$ 
by putting the boundary metric in the synchronous coordinate system
\bb\label{cosmo}
ds_{bound}^2=-d\tau^2+\frac{u(\tau)^2}{l^2} dx_{(3)}^2\ .
\ee
This simplifies comparisons between boundary vacua by
factoring in the redshift factor between the coordinate $t$ and the
boundary time variable $\tau$ uniformly in all expressions for the energy
momentum tensor we will write.
We first cut the space at a constant value $u=u_c$; 
then using~\pref{enbal}, we obtain an expression for the vev of the
$\tau \tau$ component of the energy momentum tensor (we also have
$D_a\phi^I=0$)
\bb\label{tautau}
\lk<T_{\tau \tau}\re>=\frac{6}{l} \alp_G \lk(1-h(u_c)^{1/2}\re)\simeq
3\frac{\alp_G}{l} \frac{u_0^4}{u_c^4}+O\lk[\lk(\frac{u_0}{u_c}\re)^8\re]\ .
\ee
The first term in the expansion was 
computed in~\cite{BALASTRESS} and~\cite{CO}; 
it is the contribution of a gas at temperature $T$
in the strongly interacting conformal field theory.
The higher order corrections at finite
$u_c$ become important as the temperature approaches the
UV cutoff. This phenomenon was also pointed out in the scaling of
boundary correlation functions in~\cite{SUSSHOLO} and~\cite{BALARG}. It has been
suggested that it indicates that the conformal field theory is not
enough to account for the holographic image of the bulk, and new unknown
physics, probably non-local, is needed to remedy the mismatch. 
Perhaps the same degrees of freedom are responsible for both this phenomena 
and for the presence of higher order corrections in powers of $R$
in~\pref{rsquare}. In the conventional AdS/CFT correspondence,
the boundary is taken to infinity and these effects are dropped
\footnote{Equation~\pref{tautau} for example
will get multiplied by four powers of $u_c$ to 
measure mass in the flat space at infinity; hence the first term 
remains finite when $u_c\rightarrow \infty$.}.
Looking back at equation~\pref{enbal}, we note that the boundary
metric is flat, and the conformal field theory energy is being heated by
the bulk/boundary interface. More generally, the boundary gravity will
sourced by the $K$ and $U$ terms in addition to the conformal field
theory contribution. Fixing the bulk solution and tempering with the
embedding of the boundary, one will get different four dimensional
space time and different contributions from the $K$ and $U$ terms, but 
we must still see the same conformal field theory vacuum.
Following~\cite{GUBSERRS}, let us cut the
space at $u=u_c(t)$ such that the $K$ terms in~\pref{enbal} cancel the
$U$ term; then one has, by construction, a pseudo-realistic cosmology
sourced by the conformal field theory with
\bb\label{tau2}
\lk<T_{\tau \tau}\re>=3\frac{\alp_G}{l} \frac{u_0^4}{u_c^4}\ .
\ee
This is the whole animal; no expansion in $u_0/u_c$ arises. And it is
equal to the leading term of~\pref{tautau}. And hence
the boundary theory's energy momentum tensor vev does not change.
For an arbitrary cut, the four dimensional Einstein tensor
$G_{\tau\tau}=3 \dot{u}^2/u^2\equiv 3 \alp_G H$ 
(or the Hubble parameter $H$ for the metric~\pref{cosmo})
will involve an expansion in powers of $u$ with always a term of the
form~\pref{tau2} identifying the vacuum state of the conformal field theory.
The additional terms should be given another physical interpretation
that we will study to some extent below. Yet another example for a 
boundary is given in equation~\pref{Heq} where the reader can now 
identify the now familiar
conformal field theory contribution. Equation~\pref{enbal} then
determines the vev of the boundary theory's energy momentum tensor
modulo this effect.

Let us recap our observations. In the traditional AdS/CFT correspondence,
the boundary is fixed as in~\pref{tautau} and taken to infinity. 
The holographic image to the bulk is just a conformal field theory.
When the space is sliced at finite cutoff, as in the renomalization
group picture (or equivalently the Randall-Sundrum
scenario), the holographic projection of the bulk cannot be accounted
for by the conformal field theory alone:
finite cutoff effects arise as we approach the boundary as
in~\pref{tautau} or~\pref{rsquare}; and
the extrinsic curvature of the embedding of the
boundary in the bulk introduces new contributions in the boundary theory.
We will term collectively this additional dynamics beyond the
conformal field theory as interface effects.
Fixing a bulk solution and tempering with the embedding of the boundary
does not change the dual conformal field theory, but rather alters
the boundary theory through these interface effects.
This may be interpreted as a manifestation of coordinate invariance
in the bulk in the dynamics of the boundary theory.
In some sense, it is reminiscent of the Unruh effect.
In foliating space with timelike boundaries, we are dealing however
with a somewhat
perverse manifestation of this physics, in that we are ``boosting''
the boundary in a direction which is also 
identified with renormalization group scale.
We conclude that the initial value formulation in the bulk is consistent with
the formulation required to define an effective boundary theory dual to
the bulk. In this context,
it is important to emphasize that the bulk/boundary correspondence
is to be established at the level of vacua on both sides of the duality.
We will try next to find a physical interpretation for some class of
interface effects in the boundary theory.

\section{A little cosmology}

One can play
games with equation~\pref{enbal} by adjusting the cut so as to obtain one's
favorite cosmology on the boundary, and hope to obtain
a physical understanding for the additional powers of
$u$ that will typically contribute to the Hubble parameter $G_{\tau\tau}$.
We will argue next
in favor of a scenario where these terms may be associated to
the dynamics of a probe
in the background geometry of a large number of D3 branes.

Consider the following picture\footnote{
This probe brane scenario was instigated by comments made by P. Argyres.
{\em Note added:}
After this work was complete, we also learned of~\cite{KEH} where a more
detailed analysis along this line of thought was presented. The authors
there studied dynamics of a $Dp$ brane probe in the background
geometry of a large number of $Dp'$ branes; 
they also investigated the effect of exciting
the matter sector on the brane, and argued for the resolution of the
singularity in the cosmology on the probe.}:
we are in the near horizon region of
a large number $N$ of D3 branes
at finite temperature, represented by a bulk geometry described
in the previous paragraph. The cosmological constant in the bulk is
negative and large in string units so as to keep the curvature scale
small. Another D3 brane (or a few of them) described by a U(1) (or
SU(2), or SU(3), etc.) Yang-Mills theory is freely moving in this space
in parallel orientation under the gravitational spell of the $N$ D3
branes. We also
set initial conditions such that this probe is moving away from the horizon 
with energy scale much greater than the Hawking temperature.
In a classical approximation scheme
with the back reaction of the light probe on the geometry being ignored, 
the dynamics is described
by the Dirac-Born-Infeld action. We assume the probe brane has only dynamics
given by the vevs of the scalar field corresponding to the bulk $u$ 
direction; the fluctuations of the gauge fields and other scalars are
frozen to zero by initial conditions. Obviously, we are not shooting
for a realistic scenario by making this assumption, as cosmology of interest
is to be driven by the SYM theory of small gauge group rank.
The dynamics is purely gravitational; the RR flux induces a constant shift
in the energy, and
the dilaton is constant. The trajectory of this probe in the $u-t$ plane
is obtained by minimizing the word-volume; the action is\footnote{
There is a subtlety in this description due to the infinite extent of the
probe branes. One reguralizes the dynamics by wrapping all D3 branes on 
a three torus of finite size, finds the trajectory, then takes the limit
of infinite torus size at the end.}
\bb
S_{\mbox{DBI}}\sim \int dt\ \sqrt{ h(u) \frac{u^8}{l^8} + h^{-1}(u) 
\frac{u^4}{l^4} \dot{u}^2}+1\ ;
\ee
which leads to
\bb
\frac{du}{dt}=\frac{h(u) u^2}{l^2}\lk(
1-\lk(\frac{u}{u_m}\re)^8 \frac{h(u)}{h(u_m)}\re)^{1/2}\ .
\ee
This probe samples the bulk space at progressively larger values of
$u$ before turning around and falling back into the horizon. Its maximum
reach is denoted by $u_m$ and is a constant parameter related to a 
measure of the energy of the projectile.
We may now
replace the bulk space by one with a cut just outside the trajectory
of this probe. This does not affect the dynamics in any away. It however
allows us to replace the bulk space by the boundary theory given by
equation~\pref{baction} subject to~\pref{enbal} 
using the bulk/boundary duality. 
The left hand side of equation~\pref{enbal}
then encodes the dynamical information
about the trajectory of the probe; essentially, this is the
matter sector of the probe SYM theory with the contribution
coming from the evolving vev of the $u$ field. We may then
propose that a miserable group of people confined to the surface of
the probe brane would witness a cosmology given by the induced metric
$h_{ab}$ with the Yang-Mills sector gluons frozen in their vacuum state; the
four dimensional Hubble constant is 
\bb\label{Heq}
H=\frac{u_0^4}{l^2 u^4} + \frac{u_m^8}{l^2 u^8} h(u_m) -\frac{1}{l^2}\ .
\ee
Note that we have $u_o\ll u \le u_m$. $H$ is defined by 
\bb
H\equiv \lk(\frac{\dot{u}(\tau)}{u}\re)^2\ .
\ee
The first term in~\pref{Heq} is the contribution from the conformal
field theory we saw on two other occasions earlier; 
and we conclude that the vev of the conformal field theory
energy momentum tensor is unchanged as before. The additional pieces
coming from the $K$ terms in~\pref{enbal} are to be interpreted as
contributions from the probe dynamics. We propose that this interpretation
for the meaning of the cut is a general one. The timelike
bulk/boundary interface defines a time dependent UV cutoff (with respect
to the canonical choice of time with $u=\mbox{constant}$) which should
be regarded as a $D$ dimensional dynamical effect. 
The choice of a cut conveys to the
boundary theory through~\pref{enbal}
both the vev of the energy momentum tensor of the
conformal field theory as well as the dynamics of the trajectory, all
packaged together in the extrinsic curvature term.
This idea was hinted at in~\cite{GUBSERRS} 
by adding a matter sector to~\pref{enbal}
to cancel the extrinsic curvature contribution.
The cosmology on the brane for $H$ given by~\pref{Heq} looks like
\footnote{
It is amusing that, had the trajectory been that of a point particle, 
the metric would have been
\bb
ds_{bound}^2 = -d\tau^2 + \lk(\frac{\varepsilon^2}{2} + \sqrt{
\frac{\varepsilon^4}{4}+u_0^4} \sin (\tau-\tau_0)\re) dx_{(3)}^2\ ,
\ee
where $\varepsilon$ is the energy of the probe.
The difference arises due to the fact that the space parallel to the
D3 branes warps as a function of $u$ as well; and that timelike
geodesics for spacetimes related to each other by conformal rescaling
are different.
}
\bb
ds_{bound}^2 = -d\tau^2 + \sqrt{u_0^4 + \frac{1}{2}\sqrt{
4u_m^8 h(u_m) +u_0^8} \sin \frac{4}{l}(\tau-\tau_0)} dx_{(3)}^2\ ,
\ee
This universe initially 
expands quickly as the projectile recedes away from the horizon,
then slows down reaching a maximum size, and turns around to head back for
the singularity inside the horizon to end in a spectacular crunch.
The bottom line of this discussion is
that understanding four dimensional cosmology
in such a picture involves understanding the proper dynamics of
probes in the near horizon geometry of a large number of D3 branes. 
Fluctuations on the probe must be crucial to any realistic
scenario and these 
have been ignored\footnote{We note that in the solution 
of~\cite{GUBSERRS} the
cosmology is also 
sourced by the conformal field theory, not the matter sector.}.
The suggestion is that there 
may exist a dynamical problem in the near horizon
geometry of D3 branes whose solution leads to a realistic scenario for
cosmology on the brane.

\section{Hamiltonian equations and speculations}

In the previous section, we argued that the initial value data
required to specify a bulk vacuum is consistent with the prescription
needed to define an effective boundary theory at finite cutoff. However,
given that the Hamiltonian formulation specifies two sets
of evolution equations, a set for the canonical variables, and a set for
the momenta, one may worry about the consistency of the generated flow
with our expectations regarding the behavior of a generic
quantum field theory as a function of energy scale. For example,
on the field theory side, the flow of the expectation values
of operators in the boundary theory is related to that of the couplings
since the product in the action must have a mass dimension
exactly equal to $D-1$ along the flow. This implies a relation between
the evolution equations for the scalar fields and their canonical momenta,
which we already know exists in the Hamiltonian formulation.
That Hamiltonian flow, with its property to preserve phase space volume,
may realize the structural form of
renormalization group flow equations in this sense was
first observed in~\cite{DOLAN}. More generally,
perhaps the interpretation may be that 
the full form of the first order Hamiltonian differential 
equations do not present any puzzles with regards to what we know about
quantum field theories, but point to new characteristics of 
theories with dual gravitational descriptions\footnote{
This possibility was pointed out by P. Argyres.}. In this section,
we make a mostly unsuccessful attempt to address these issues.
We present a set of representative formulas for readers to stare at
and possibly get inspired by.

Let us systematically look at the first order differential 
equations arising in the Hamiltonian formalism.
The two canonical variables $h_{ab}$ and $\phi^I$ satisfy the equations
\bb\label{dothdotphi}
\dot{h}_{ab}=2 \lk(\tau_{ab}-h_{ab} \frac{\tau_c^c}{D-2}\re)\ ,\ \ \ 
\dot{\phi}^I= - G^{IJ} \Omega_J\ .
\ee
The dot stands for differentiation with
respect to the coordinate $u$ normal to the boundary; the reader is
referred to the appendix for the details. We note that, the gauge
in which we have written these equations
corresponds to the choice of Gaussian normal coordinates where $u$ 
is the parameter along spacelike geodesics projected from the boundary.
Bousso's criterion for holography is a statement about the rate
of convergence of null geodesics projected from a boundary. It was
shown in~\cite{CFUNCTION} that this statement can be connected to the
monotonicity of the c-function of the boundary theory. In general,
it is natural to expect that one needs to pick an ``arrow'' in Hamiltonian
flow to map the flow equations onto renormalization group given that
the latter naturally runs from higher energies to lower.
In the context of a timelike boundary, noting that the normals to the
foliations are tangents to spacelike geodesics, we may propose
a statement of the same form as~\cite{BOUSSOCONJ,BOUSSOCONJ2}; we require the
trace of the extrinsic curvature to be non-positive $K\le 0$. This may follow
from the criterion with respect to
null geodesics, but we will not worry
about the connection, and instead propose a stronger version;
we will require that the eigenvalues of
$K_{ab}$ are non-positive if the bulk is to have 
a holographic image on the boundary. Since $K_{ab}$ is the Lie
derivative of the induced metric $h_{ab}$ along the flow lines (see
appendix), the physical content of this statement is that one needs
the induced boundary metric to warp in a monotonically decreasing
manner as one moves deep into the bulk space. The scenario
in~\cite{dBVV}
assumed that the warping is isotropic. More importantly, this new statement
regarding monotonicity of the flow of the metric
allows us to implicitly invert the $u$ dependence of all the variables
to one with respect to the boundary metric
\bb
\frac{d}{du}=\int \dot{h}_{ab} \frac{\delta}{\delta h_{ab}}\ .
\ee
Let us use this statement on $\dot{\phi}^I$ with equations~\pref{dothdotphi}
and~\pref{baction}; we then get 
\bb\label{beta0}
\int h_{ab} \frac{\delta \phi^I}{\delta h_{ab}}=
- \frac{\alp_G}{\alp_M} \frac{D-2}{U}
G^{IJ} \del_J U\ .
\ee
A natural definition for the beta functions of the boundary theory in
this context is
\bb\label{beta}
\beta^I\equiv 2\int h_{ab} \frac{\delta \phi^I}{\delta h_{ab}}\ .
\ee
The factor of two comes from the fact that the metric scales with two
powers of length. 
Equations~\pref{beta0},~\pref{beta},~\pref{tcc} and~\pref{tcc2}
allow us to determine the correct normalization for
the energy momentum tensor of the boundary theory
\bb
2 \lk< t_{ab}\re> = \lk< T_{ab} \re>\ ,
\ee
which we already made use of in the text.
We emphasize that
our definition for $T_{ab}$ is part of the whole animal, the part that
acquires only an anomaly of the $\beta^I \lk<O_I\re>$ form. In particular,
gravitational anomaly terms have been factored away.

The other two sets of equations are those specifying the evolution of
the canonical momenta
\bbb\label{dotomega}
\dot{\Omega}_I&=&\del_I V -\frac{1}{2} \del_I G_{KL} D_c \phi^K D^c \phi^L
+\del_K G_{IJ} D_c \phi^K D^c \phi^J+G_{IJ} D_c D^c \phi^J \nonumber \\
&+&\frac{1}{2} \del_I G^{KL} \Omega_K \Omega_L
+\frac{1}{D-2} \Omega_I \tau_c^c\ ;
\eee
\bbb\label{dottau}
\dot{\tau}_{ab}&=&G_{ab}-h_{ab} \Lambda -\frac{1}{2} h_{ab} \lk(
\tau_{cd} \tau^{cd} - \frac{\lk(\tau_c^c\re)^2}{D-2}\re)
+ 2 \lk(\tau_{ac}\tau^c_b
-\frac{1}{2} \frac{\tau_c^c \tau_{ab}}{D-2}\re) \nonumber \\
&+&\frac{1}{2} h_{ab} \lk( \frac{1}{2} G_{IJ} D_c \phi^I D^c \phi^J
-\frac{1}{2} G^{IJ} \Omega_I \Omega_J-V(\phi)\re) \nonumber \\
&-&\frac{1}{2} G_{IJ} D_a \phi^I D_b \phi^J\ .
\eee
Trading derivatives with respect to $u$ for derivatives with
respect to $h_{ab}$ in~\pref{dotomega}, and after somewhat lengthy
manipulations, one gets
\bb\label{torel}
\int_x \beta^J \lk<\OO_I(x) \OO_J(0)\re>
+\lk(\del_J \beta_I\re) G^{KJ} \lk<\OO_K\re>
+\lk(\del_I G^{KL}\re) \beta_K \lk<\OO_L\re>
+(D-1) \lk< \OO_I\re>=0\ .
\ee
In handling these expressions, the process is somewhat tedious 
due to the fact that, at various stages of the expansion,
one gets relations or derivatives of relations obtained independently
from other Hamiltonian equations or the constraint equation. So, one
gets a sea of useless information, buried within it a few interesting 
statements such as~\pref{torel}. The structure of the latter is suggestive;
if one could replace $\beta^I \OO_I$ with $T_c^c$ (which one cannot
except when the vevs are taken), and if the boundary theory
is conformal, this is just the Ward identity for the dilation current
and one identifies the anomalous dimension of the operator $\OO_I$ as
\bb
\Delta_I=\lk(\del_J \beta_I\re) G^{KJ} +\lk(\del_I G^{KL}\re) \beta_K+D-1\ ,
\ee
which is exactly as expected from an effective $D-1$ dimensional
field theory on the boundary. Note that the scalars are to be identified
with the dimensionless couplings.
The middle piece is non-zero when the kinetic
term for the scalars in the bulk is dressed by additional powers of the
scalar fields (as in non-linear sigma models).
Equation~\pref{torel} is however some sort of integral of these statements
and it is not apparent to us how to extract the suggestive information
about the operator's dimension from this expression.
We may however speculate that it appears structurally
that the evolution of the $\phi^I$'s and the $\Omega_I$'s
``know'' about each other in the manner required by a $D-1$
dimensional quantum field theory.
We also have not been able to pinpoint to a
general principle assuring consistency of the content of the
Hamiltonian equations with renormalization group flow in general. 

Let us comment briefly 
on some of the other physics
in equations~\pref{dotomega} and~\pref{dottau}. 
If we assume monotonicity of
the couplings as well (thus restricting to a class of operators 
along a particular flow line), 
we can invert the $u$ derivative in $\dot{\Omega}_I$ for one for $\phi^J$
and obtain an equation involving the correlators of two operators
\bbb\label{nasty}
\int_x \lk. G^{JK} M_{KA}\re|_x \lk<\OO_J(x) \OO_I(0)\re>&=&
\del_I M_{AJ} G^{JK} \lk<\OO_K\re>
-\del_I G^{KL} M_{AL} \lk<\OO_K\re> \nonumber \\
&-&\frac{\alp_M}{2(D-2) \alp_G} \beta^J M_{AI} \lk<\OO_J\re>
+\frac{\alp_M}{\alp_G} \del_A \Phi \lk<\OO_I\re>\ .
\eee
We remind the reader that $M_{IJ}$ is the metric over coupling space.
Again this can be argued to have a suggestive form, but we are not
able to extract the OPE coefficients from this expression.
A myriad of other relations may also be written between the correlation
functions, and we write this one as a representative for the
form these equations generically take.
Playing the same game with the $\dot{\tau}_{ab}$ equation, one gets
relations for the correlators for the energy momentum tensor
with itself and all the operators in the theory.

One may hope that the Hamiltonian evolution equations can be used to
systematically {\em construct} an effective boundary theory by
extracting the expectation values of all the correlators of the
theory from such equations; for higher point correlators, one looks at
higher derivatives of the canonical variables and trades the $u$ derivatives
with variations with respect to the other variables. The effective action
piece in~\pref{baction} then spits out higher point correlation functions.
This is a laborious but straightforward procedure; at every stage,
one gets pieces of new physics about the boundary.
It is in this sense that one may hope to systematically construct a $D-1$
dimensional boundary theory dual to a $D$ dimensional gravitational bulk.
One would even hope to say then that this constitutes a proof that 
the equations of
motion of any gravitational theory are equivalent to defining a $D-1$
dimensional quantum field theory. The equations satisfied by 
the correlators, such as~\pref{nasty}, may also be viewed as restrictions
on the class of theories which admit gravitational duals.

We end the discussion in this section by applying some of the previous 
equations to Dp branes.
We are looking for the beta function of the Yang-Mills coupling as encoded
in the extremal geometries.
The trace of the energy momentum tensor then becomes
\bb
\lk<T^c_c\re>_{h_{ab},f}=\frac{p}{2} \frac{(p-3)(7-p)}{(9-p)}
\del_\phi f\ \lk<\LL_{SYM}\re>_{h_{ab},f}\ ,
\ee
where 
\bb
f(\phi)\equiv - N \lk( N e^\phi\re)^{2\frac{p-9}{p(7-p)}}\ .
\ee
Rescalings this expression to the frame where the Yang-Mills is
coupled to a Minkowski metric, most factors cancel and one obtains 
the simple result\footnote{
In particular, the transformations are
\bb
\lk<\LL_{SYM}\re>_{h_{ab},f}\rightarrow \Omega^{-4}
\lk<\LL_{SYM}\re>_{h_{ab}^{dual},\Omega^{p-3} f}\ ,\ \ \ 
\lk<T_{ab}\re>_{h_{ab},f}\rightarrow \Omega^{-p-1}
\lk<T_{ab}\re>_{h_{ab}^{dual},\Omega^{p-3} f}\ .
\ee
}
\bb\label{bym}
\beta_{\geff^2}=(p-3)\geff^2 + \cdots \ .
\ee
This is precisely the evolution of the dimensionless Yang-Mills
coupling according to its classical dimension. Higher order quantum
corrections are then encoded in string theoretical corrections to
the bulk action. 
We saw earlier that $R^2$ terms in the
bulk will readily lead to corrections to the gravitational
coupling, and hence one would expect to the c-function. The
beta functions being gradients of the c-function~\cite{dBVV}, we
expect an expansion in negative powers of the Yang-Mills effective 
coupling corresponding to $\alpha'$ corrections in the bulk.
Furthermore, it can be checked that the beta function 
we obtained at this order is the gradient of the
c-function~\pref{cfunc} with the metric over the one dimensional
coupling space taken as 
$M_{\geff^2}$ in~\pref{mgeff}
\bb
\beta_{\geff^2} = \frac{\Omega_{8-p}}{(2\pi)^7} \frac{9-p}{p (5-p)}
M_{\geff^2}^{-1} \del_{\geff^2} c(\geff^2)\ .
\ee

\section{Appendix A: Conventions and notation}

We review in this appendix the Hamiltonian formalism used in the text.
We note in particular important flips of signs between our formulas 
and the ones found in standard textbooks due to the fact that the boundary
is timelike. Otherwise, we follow closely~\cite{WALD}. 

The bulk action is given by
\bb\label{action2}
S=\alp_G \int_D \lk( R+2\Lambda\re)
+ 2\alp_G \int_{D-1} K
+ \alp_M \int_D V(\phi)
-\frac{1}{2} G_{IJ} \nabla_\mu \phi^I \nabla^\mu \phi^J\ . 
\ee
An arbitrary foliation is chosen and the space is cut along a timelike
surface of this foliation generating a $D-1$ dimensional boundary;
the foliations are defined by the scalar function $u=\mbox{constant}$;
and the vector $u^a$ is chosen such that $u^a\nabla_a u =+1$.
The spacelike {\em inward pointing}
normals to the foliations, $n_a$, are normalized $g_{ab}\ n^a n^b=+1$
and the boundary metric $h_{ab}$ is given by $g_{ab}=h_{ab} + n_a n_b$.
The extrinsic curvature is 
$K_{ab}\equiv h_a^c \nabla_c n_b$;
this tensor is symmetric by virtue of the surface orthogonality 
of the normals and it is transverse on both of its indices by construction.
We define the lapse function $N$ and the lapse vector $N^a$ by
$u^a\equiv N^a - N n^a$.  The metric then takes the form
\bb
ds^2=\lk(N^2+N_a N^a\re) du^2+ 2 N_a du dx^a+h_{ab} dx^a dx^b\ .
\ee

Splitting the action along the foliations yields
\bb
\LL_{tot}=-\alp_G \sqrt{h} N \lk( R^{\{D-1\}}+K^2-K_{ab} K^{ab}+2\Lambda\re)
-\alp_M \sqrt{h} N \LL_M\ .
\ee
Here $R^{\{D-1\}}$ is the $D-1$ dimensional Ricci scalar; in the text
and hereafter,
we drop the $D-1$ superscript to avoid cluttering formulas.
We have
\bb
K_{ab}=\frac{1}{2} \LL_n h_{ab}\ ,\Rightarrow \ 
2 N\ K^{ab}=\dot{h}^{ab} + D^a N^b + D^b N^a\ .
\ee
The canonical momenta are defined by
\bb
\tau_{ab}\equiv \frac{1}{\alp_G \sqrt{h}}
\frac{\delta \LL_{tot}}{\delta \dot{h}^{ab}}=
\lk(K_{ab} - K h_{ab}\re) \ ,\ \ \ 
\Omega_I\equiv \frac{1}{\alp_M \sqrt{h}}
\frac{\delta \LL_{tot}}{\delta \dot{\phi}^I}=
\frac{1}{N} G_{IJ} \lk( \dot{\phi}^J-N^a D_a \phi^J\re)\ .
\ee
The Hamiltonian then becomes
\bbb
\HH&=&\alp_G \sqrt{h} N \lk(R+2\Lambda\re)
+\alp_G \sqrt{h} N \lk( \tau_{ab} \tau^{ab} -\frac{\lk(\tau^c_c\re)^2}{D-2}\re)
\nonumber \\
&-&\alp_G \sqrt{h} \tau_{ab} \lk(D^a N^b+D^b N^a\re)
+ \alp_M \sqrt{h}\ \Omega_J N^a D_a \phi^J \nonumber \\
&-&\alp_M \sqrt{h} N \lk( \frac{1}{2} G_{IJ} D_c\phi^I D^c\phi^J
-\frac{1}{2} G^{IJ} \Omega_I \Omega_J-V(\phi)\re)\ .
\eee
The constraint equations result from varying $N$ and $N^a$; they are given 
in~\pref{cons} and \pref{encons} respectively.

In the text, we choose throughout the gauge $N^a=0$ and
$N=-1$ corresponding to Gaussian normal coordinates.
The normals are then tangents to spacelike geodesics and $u$ is
the parameter along these geodesics.
The equations of motion are given by~\pref{dothdotphi},
\pref{dotomega} and~\pref{dottau} with $N^a=0$, $N=-1$.

\paragraph{\bf Acknowledgments:}
I am grateful to P. Argyres, 
M. Moriconi, H. Tye and I. Wasserman for discussions.
This work was supported by NSF grant 9513717.

\providecommand{\href}[2]{#2}\begingroup\raggedright\endgroup

\end{document}